\documentstyle[12pt,epsfig,amssymb]{article}
\voffset     = - 0.50 in
\hoffset     = - 0.20 in
\begin{document}
\thispagestyle{empty}

%{\large\bf
%\hfill
%\begin{tabular}{r}
%ATLAS Internal Note                \\
%TILE-CAL-NO-72                     \\
%\today                             \\
%\end{tabular}
%}

\vspace*{\fill}

{\Large\bf
\begin{center}
Electron Response and $e / h$ Ratio of  
Iron-Scintillator Hadron  Prototype Calorimeter  
with Longitudinal Tile Configuration \\
\end{center}
}

\vspace*{\fill}

\begin{center}

{\Large\bf
Y.A.~Kulchitsky, \\
}
{\it
Institute of Physics, Academy of Sciences, Minsk, Belarus \\
\& JINR, Dubna, Russia
}

%\bigskip

{\Large\bf
J.A.~Budagov, V.B.~Vinogradov \\
}
{\it
JINR, Dubna, Russia
}

%\bigskip

{\Large\bf
M.~Nessi \\
}

{\it
CERN, Geneva, Switzerland
}

%\bigskip

{\Large\bf
A.A.~Bogush \\
}

{\it
Institute of Physics, Academy of Sciences, Minsk, Belarus
}

%\bigskip

{\Large\bf
V.V.~Arkadov \\
}
{\it
MEPI, Moscow {\it \&} JINR, Dubna, Russia
}

%\bigskip

{\Large\bf
G.V.~Karapetian \\
}
{\it
YerPI, Yerevan, Armenia {\it \&}
CERN, Geneva, Switzerland
}

\end{center}

\vspace*{\fill}

\begin{abstract}
The detailed information about electron response, electron energy resolution
and $e / h$ ratio as a function of incident energy $E$, impact point $Z$
and incidence
angle $\Theta$ of iron-scintillator hadron prototype calorimeter with
longitudinal tile configuration is presented.
These results are based on
electron and pion beams data of $E = 20,\ 50,\ 100,\ 150,\ 300\  GeV$ at
$\Theta = 10^{o},\ 20^{o},\ 30^{o}$,
which were obtained during test beam period in July 1995.
The obtained calibration constant is used for muon response converting
from $pC$ to $GeV$.
The results are compared with existing experimental data and with some
Monte Carlo calculations.
For some $E,\ \Theta,\ Z$ values the local compensation ($e / h = 1$)
is observed.
\end{abstract}

%\vspace*{\fill}

\newpage

{
%\large

%\tableofcontents

\section{Introduction}

The ATLAS Collaboration proposes to build a general-purpose pp detector
which is designed to exploit the full discovery potential of the CERN's
Large Hadron Collider (LHC), a super-conducting ring to provide
proton -- proton collisions around 14 TeV \cite{atcol}.
LHC will open up new physics horizons, probing interactions between proton
constituents at the 1 TeV level, where new behavior is expected to reveal
key insights into the underlying mechanisms of Nature \cite{lhcnews}.

The bulk of the hadronic calorimetry in the ATLAS detector is provided
by a large (11 m in length, 8.5 m in outer diameter, 2 m in thickness,
10000 readout channels)
scintillating tile hadronic barrel calorimeter.

The technology for this calorimeter is based on a sampling technique
using steel absorber material and scintillating plates readout by
wavelength shifting fibres.
An innovative feature of this design is the orientation of the scintillating
tiles which are placed in planes perpendicular to the colliding beams
staggered in depth \cite{gild-91} (Fig.~\ref{f2-28tp}).
This geometry combines good performance and a simple and cost effective
assembly procedure \cite{ariz-94}.

In order to test this concept five module prototype of a calorimeter
was built and exposed to high energy pion, electron and muon beams at the
CERN Super Proton Synchrotron.
Results on response, energy resolution and linearity were obtained which
confirmed the righteousness of the proposed concept
\cite{atcol}, \cite{ariz-94}, \cite{berger}.
%But as to electron response and especially $e / h$ ratios the data were unsufficient.

The aim of this work is to obtain the detailed information about the
electron response of this calorimeter and $e / h$ ratios.
This information is based on data which were obtained during test beam
period in July 1995.
Some results about the electron response are obtained in
\cite{david} and \cite{efthym}.

\section{The Prototype Calorimeter}

The prototype calorimeter is composed of five sector modules, each
spanning $2 \pi / 64$ in azimuth, 100 cm in the axial ($Z$) direction,
180 cm in the radial direction (about 9 interaction lengths),
and with  a front face of 100 $\times$ 20 cm$^2$ \cite{berger}.
The iron structure of each module consists of 57 repeated "periods".
Each period is 18~mm thick and consists of four layers (Fig.~\ref{f2-29tp}).
The first and third layers are formed by large trapezoidal steel plates
(master plates), 5 mm thick and spanning the full radial dimension of the
module.
In the second and fourth layers, smaller trapezoidal steel plates
(spacer plates) and scintillator tiles alternate along the radial direction.
The spacer plates are 4 mm thick and of 11 different sizes.
Scintillator tiles are 3 mm thickness.
The iron to scintillator ratio is 4.67:1 by volume.

Radially oriented WLS fibres collect light from the tiles at both of
their open edges and bring it to photo-multipliers (PMTs) at the periphery
of the calorimeter.
Each PMT views a specific group of tiles, through the corresponding
bundle of fibres. With this readout scheme three-dimensional segmentation
is immediately obtained.

Tiles of 18 different shapes all have the same radial dimen\-si\-ons (10 cm).
The prototype calorimeter is radially segmented into four depth segments by
grouping fibres from different tiles.
Proceeding outward in radius, the three smallest tiles,
$1  \div  3$ are grouped into section 1,
$4  \div  7$ into section 2,
$8  \div 12$ into section 3 and
$13 \div 18$ into section 4.
The readout cell width in $Z$  direction is about 20 cm.

\section{Test Beam Layout}

The five modules have been positioned on a scanning table, able to allow
high precision movements along any direction.
Upstream of the calorimeter, a trigger counter telescope was installed,
defining a beam spot of 2~cm diameter.
Two delay-line wire chambers, each with $Z$, $Y$  readout,
allowed to reconstruct the impact point of beam particles on the
calorimeter face to better than $\pm$~1~mm \cite{ariz-94}.
A helium \v{C}erenkov threshold counter was used to tag
$\pi$-mesons and electrons for $E$ = 20 $GeV$.
A large scintillator wall covering about 1~$m^2$ of surface has been
placed on the side and on the back of the calorimeter to quantify back
and side leakage.

\section{Data Taking and Event Selection}

Data were taken with electron and pion beam of 20, 50, 100, 150, 300 $GeV$
at $\Theta = 10^{o} , 20^{o} , 30^{o}$ and
$Z  = -~8 \div -~36$ cm.
Number of runs about 60, number of events is approximately a half of million.
The treatment was carried out using program TILEMON \cite{atmon}.

As a result we have for each event 200 values of charges $Q_{ijkl}$
from PMT properly calibrated \cite{berger} with pedestal subtracted.
Here indexes $i,\ j,\ k,\ l$ mean:
$i = 1, \ldots, 5$ is the raw number, $j = 1, \ldots, 5$ is the module number,
$k = 1, \ldots, 4$ is the depth number and $l = 1, 2$ is the PMT number.

The Table~\ref{Tb1} represents the volume of analyzed information.
To separate electrons from pions the following criteria were proposed
which are given in Table~\ref{Tb2}.
The cuts 1 and 2 remove beam halo.
The cut 3 removes muons and nonsingle-track events.
The cuts 4 and 5 carry out electron-pion separation
(Fig.~\ref{fig:f4} $\div$ Fig.~\ref{fig:f6}).
The cut 4 is connected with \v{C}erenkov counter amplitude ($S_{Cer}$).
Cut 5 is the relative shower energy deposition in the first two
calorimeter depths, where the indexes $i$ and $k$ in $Q_{ijkl}$ determine
the regions of electromagnetic shower development and
\begin{equation}
\label{ci}
C_{i} = \sum_{selected\  i}\  \sum_{j = 3} \sum_{k = 1}^{2}
\sum_{l = 1}^{2} Q_{ijkl} / E,
\end{equation}

\begin{equation}
\label{en}
E = \sum_{ijkl} Q_{ijkl}.
\end{equation}
The values $C_i$ depend on a particle's entry angle $\Theta$.

The basis for electron-hadron separation is the different longitudinal
energy deposition for electrons and hadrons.
A particle traversing the first two calorimeter depths at $10^{o} \div 30^{o}$ angles crosses
$55 \div 63$ cm of iron.
It corresponds to $31 \div 36$ radiation length or $3.3 \div 3.8$ nuclear
interaction ones.
The amount of energy deposited is equal to approximately 100\% for the
electromagnetic shower and about half for hadronic one \cite{amaldi}.

Fig.~\ref{fig:f4} shows the distribution of electron and pion events as a
function of $E$ and the value of \v{C}erenkov counter signal.
Two group of events are observed.

Figures~\ref{fig:f5a} $\div$ \ref{fig:f5c}  show the distribution of the
electron and pion events for
$E = 20\ GeV$ at $\Theta = 10^{o}$ as a function of $C_i$
for electrons (Fig.~\ref{fig:f5a}) and for pions (Fig.~\ref{fig:f5b}) tagged
by \v{C}erenkov counter and for all events (Fig.~\ref{fig:f5c}).

The values $\sigma$ for each energy were obtained by Gaussian fit of spectra
region around 1 where pronounced peaks are observed (Fig.~\ref{fig:f6}).

From these Figures some estimations of pion and electron contaminations
were obtained.
At 20 $GeV$ electron contamination in pion region does not exceed the 3\%
level and pion contamination in the electron region is less 6\%.
These contaminations considerably decrease with energy increasing.

Fig.~\ref{fig:f6} shows the distribution of all events as a function of $C_i$
for $E = 300\ GeV$\ at $\Theta = 30^{o}$.
At this energy pion contamination in electron region does not exceed the 1\%
level.

\section{Results}

\subsection{Electrons}

\subsubsection{Response}

As to electron response our calorimeter is very complicated object.
It may be imagined as a continuous set of calorimeters with the
variable absorber and scintillator thicknesses
(from $t$ = 81 to 28 mm and from $s$ = 17 to 6 mm
for $10^{o} \leq \Theta \leq 30^{o}$),
where $t$ and $s$ are the thicknesses of absorber and scintillator
respectively.

Therefore an electron response ($R = E / E_{beam}$) is rather complicated
function of $E_{beam}$, $\Theta$ and $Z$.
Energy spectrum for given run (beam has the transversal spread
$\pm 10\ mm$) is non-Gaussian (Fig.~\ref{fig:f07}), but it becomes Gaussian
for given $E,\ \Theta,\ Z$ values.

The normalized electron response  for $E = 20\ GeV$ at $\Theta = 10^o$
is shown in Fig.~\ref{fig:f7}
as a function of impact point $Z$ coordinate.
One can see the clear periodical structure of the response with 18~mm period.

The mean over two 18 mm periods responses $R$
and relative spreads ($\Delta R / R = (R_{max} - R_{min})/R_{mean}$)
as a function of $E$ and $\Theta$ are
presented in Table~\ref{Tb3}.
The relative spread
decreases with energy increasing from 16\% at  $E = 20\ GeV,\ \Theta = 10^{o}$
to 4\% at $E = 300\ GeV,\ \Theta = 30^{o}$.
For $\Theta = 10^o$ the significant relative spread decreasing with energy
increasing is observed.
For $\Theta = 20^o$ and $30^o$ spreads decrease more slowly.
In all energies the spread decreasing with the angle increasing is observed.
The mean value $R$ for all angles and energies is
$5.69 \pm 0.04\ (stat.)\ \pm 0.15\ (syst.)\ pC/GeV$.
As a systematic error the estimate of r.m.s. of distribution of values
$R$ is used.
In \cite{henriq} the experimental data on muon response have been converted
from $pC$ to $GeV$ using this calibration constant.

We attempted to explain the electron response as a function of $Z$ coordinate
calculating the total number of shower electrons (positrons)
crossing scintillator tiles using the shower curve
(the number of particles in the shower $N_{e}$ as a function of the
longitudinal shower development) which is given in \cite{abshire}.
These calculations were performed for
all energies and angles for trajectories entering into four
different elements of calorimeter periodic structure ---
spacer, master, tile, master (Fig.~\ref{f2-29tp}).
The results for $E = 20\ GeV$ at $\Theta = 10^o$
normalized and attached to the experimental response at
$Z = - 88.5\ mm$ (Fig.~\ref{fig:f7}) are given in Table~\ref{Tb05}
together with the experimental data.
Such simple calculations are in agreement with data at
$E = 20\ GeV$ but do not reproduce spread decreasing with energy
increasing.
The latter is connect with increasing the shower lateral spread
with energy increasing.

\subsubsection{Energy Resolution}

For all analyzed electron data the energy resolutions for different
values $E,\ \Theta,\ Z$ were obtained.
As an example, Fig.~\ref{fig:f11} shows the energy resolution  as
a function of $Z$ for $E = 20\ GeV$ at $\Theta = 10^{o}$.
Data reveal some structures in the $Z$ dependence but with smaller
spread than the response one.
The energy resolutions averaged over two 18 mm $Z$ period
are shown in Fig.~\ref{fig:f12} as a function of $1 / \sqrt{E}$ and presented in Table~\ref{Tb5}.
The calculations of energy resolution values were performed without restricting on the
area around the peak value.

Fits of these data by the formulae
\begin{equation}
\label{ls}
\frac{\sigma}{E}=\frac{a}{\sqrt{E}} + b,
\end{equation}

\begin{equation}
\label{qs}
\frac{\sigma}{E}=\frac{a}{\sqrt{E}}\oplus b,
\end{equation}
are given in Table~\ref{Tb6}.
These two expressions describe the data satisfactorily.
The values of statistical term  $a$ about the same for these cases,
and the values of constant term  $b$ about three times is greater
in quadratic summing (\ref{qs}).
The statistical term $a$ of the resolution decreases from
$\approx$ 65\% (for 10$^o$) to $\approx$ 35\% (for 30$^o$).
%The constant term in (\ref{qs}) do not agree for 10$^o$ and agree for 20$^o$ and 30$^o$.

The energy resolution of the 10$^o$ data is significantly worse than for
20$^o$ and 30$^o$ ones.
Consideration of the 10$^o$ trajectories traversing the calorimeter
showed that in this case on the way of shower the large iron
ununirformities exist.
For example, for the trajectory entering in to the master plate
the next structure
\begin{equation}
\label{str}
81~Fe-Sc-29~Fe-Sc-132~Fe-Sc-23~Fe-Sc-29~Fe-Sc-81~Fe-Sc
\end{equation}
is observed (thickness of $Fe$ in mm, thickness of scintillator $Sc$ is equal to
17~mm).
At the same time due to the relatively short electromagnetic shower length
only few scintillator tiles effectively works.
For example, at 20 $GeV$ 80\% shower particles cross only two
tiles.
All this result in large energy deposition fluctuations.

From some point of view, our calorimeter may be considered as a calorimeter
with variable sampling steps (\ref{str}).
In \cite{prabha} the optimization of electromagnetic
calorimeter at variable sampling steps have been studied.
It was shown, that optimization of the sampling layers (samplers) positions give
significantly better energy resolution with compare to a conventional,
constant sampling one.
In our case, probably, on the contrary
positions of samplers have not lead to the electron energy resolution improvement.
It is to be noted that it concerns only the energy
resolution for electrons but for this the calorimeter was not designed.

We compared our results on energy resolution with parametrization
from \cite{delpe}:
\begin{equation}
\label{dpe}
\frac{\sigma}{E} = \frac{a}{\sqrt{E}} =
\frac{{\sigma}_{o}}{\sqrt{E}} \cdot
{\Biggl( \frac{t}{X_t} \Biggr) }^{\gamma}
\cdot
{\Biggl( \frac{s}{X_s} \Biggr)}^{\delta},
\end{equation}
where ${\sigma}_{o} = 6.33\ \% \cdot \sqrt{GeV}$, $\gamma$ = 0.62,
$\delta$ = 0.21 are the parameters,
$X_t$ and $X_s$ are the radiation lengths of iron and scintillator
respectively.
In our case the values of $t$ and $s$ are equal to:
$t = 14~mm / \sin \Theta $, $s = 3~mm / \sin \Theta$.
This formula is purely empirical and the parameters
${\sigma}_{o}, \gamma , \delta$ were determined by fitting Monte Carlo data.

The results of calculations are given in Table~\ref{Tb11}.
As can be seen from this Table
the energy resolutions calculated by formula (\ref{dpe}) are more accurate
than experimental ones
but it is necessary to note that there is the non-adequateness of
our calorimeter in comparison with Monte Carlo calculations
(non-uniformity along direction of the shower development,
non normal incidence).
Besides, the given in \cite{delpe}
values of ${\sigma}_{o}, \gamma , \delta$ are not quite adequate for our 10$^o$
and 20$^o$ because these parameters were determined for the
iron -- scintillator with $s \leq 10$ mm and $t \leq 35$ mm.

\subsection{Pion Response}

The energy spectra for all studied energies and angles were obtained.
As an example, the observed energy spectrum for $20\ GeV$\ at $20^{o}$
is shown in Fig.~\ref{fig:f22}.
The spectrum displays small or no tails.
Above 150 $GeV$ some low energy tail appears due to longitudinal leakage of
the hadronic shower.

Pion data don't reveal any structure (Fig.~\ref{fig:f21})
with the exception of data at $E = 300\ GeV$.
%(Fig.~\ref{fig:pi})
This ununiformity was noted previously by M.~Bosman \cite{bosman}.
This effect may be related with the energy leakage.

\subsection{$e / h$ Ratio}

The responses obtained for $e$ and $\pi$ give the possibility to extract
$e / h$ value, the ratio of the calorimeter responses to the
electromagnetic ($e$) and non-electromagnetic (purely hadronic)
components of hadron showers.
As it is known the value $e/h \not= 1$ causes deviation from linearity in
the hadronic response versus energy, besides broadening the energy
resolution \cite{wigmans}.
The value of the $e / h$ ratio as shown in \cite{rwig} depends on a
big number of factors,
among them, the thickness of the passive layers, the thickness of
the active layers, the sampling fraction.
The $e / h$ ratio of a sampling calorimeter with an iron -- scintillator
ratio less than 20 is expected \cite{wigmans} to be $> 1$ for the
conventional orientation of tiles with respect to incident hadrons.

In our case the electron -- pion ratios reveal complicated structures
$e / \pi  = f( E, \Theta, Z)$.
As an example, Fig.~\ref{fig:f13} shows
$e/ \pi$ ratio as a function of $Z$ coordinate
for $E = 20\ GeV$ at $\Theta = 10^{o}$.

For some $Z$ points the local compensation
(the equalization of the electromagnetic and hadronic signals,
$e/ \pi = e / h = 1$) is observed.
This is the first experimental observation of compensation in
iron-scintillator calorimeters.
Compensation in these calorimeters had been predicted in \cite{wigmans}
but in considerably greater value of
$R_{d} = t / s$
(much more  10)
%($\simeq 80$)
than our $R_{d} = 4.7$ value.
Reconsideration has led Wigmans (\cite{wigmans}) to the conclusion that for
thick absorber plates $e / h$ value depends also on the plates thickness
and not only on $R_d$ and $e / h$ is predicted to become 1 for plates of
$\approx 110$ mm.
In our case the iron thickness is equal to $\approx 80$ mm.

The $e/ \pi$ ratios averaged over two 18 mm period, are given in
Table~\ref{Tb7} and shown in Fig.~\ref{fig:f14} as a function of
beam energy.

The $e / h$ ratios were extracted from these data by the
formula \cite{rwig}:
\begin{equation}
\label{wig}
e/ \pi = \frac{e/h} {1 + (e/h - 1) \cdot 0.11 \cdot lnE}.
\end{equation}

Fitted $e / h$ values are given in Table~\ref{Tb8}  with the other existing
experimental data for iron-scintillator calorimeters
together with the corresponding values of thickness of the iron absorber
($t$), thickness of the readout scintillator layers ($s$) and
the ratio $R_{d} = t / s$ (Fig.~\ref{fig:f17}).

The $e /h$ values generally decrease with increasing of iron thickness
(Fig.~\ref{fig:f15}).
Besides, some data, \cite{antipov} and \cite{holder}, fall out from the
general picture.
The result of \cite{antipov} may be discarded because in this
work as the authors have written
the $e / \pi$ signal ratio is roughly estimated and only at 25 $GeV$.

At the same time the considerable disagreement between different
Monte Carlo calculations \cite{wigmans}, \cite{gabriel} and experimental data
is observed.

Accordingly \cite{wigmans} the reason of $e / h$ decrease is because
the hadron signal ($h$)
increases owing to more neutrons released in hadronic shower
with iron thickness increasing
while electron signal ($e$) decreases owing to more
absorption of the electromagnetic shower.

Fig.\ref{fig:f16} shows the $e / h$ ratios as a function
of scintillator thickness.
We observe a decreasing $e / h$ value  in comparison with increasing
scintillator thickness.
Besides, it is seen that the extrapolation of our results and data
from \cite{ariz-94}, \cite{bohmer}
by smooth curve is compatible with Monte Carlo calculations \cite{gabriel}.
As can be seen from Fig.~\ref{fig:f15} and \ref{fig:f16} there is
the clear correlation between  the $e / h$ decreasing and the simultaneous
$t$ and $s$ thicknesses increasing.
This is the first experimental observation of such $e / h$ behavior.

\section{Conclusions}

We have investigated the various properties of hadron tile calorimeter
with respect to electrons.
The detailed information about response, energy resolution,
$e / \pi $-ratio as a functions of incident energy, impact point
and angle is obtained.

\bigskip

Some results are following:
\begin{itemize}
\item
The significant variation of electron response ($3 \div 16$ \%) as a
function of $Z$ coordinate with 18 mm periodic dependence is observed.
The mean value of electron response for all energies and angles
is $R = 5.69 \pm 0.04\ \pm 0.15\ pC/GeV$.
\item
The energy resolution data are well fitted by formulae (\ref{ls})
and (\ref{qs}).
The statistical term $a$ of the resolution decreases from
$\approx$ 65 \% (for 10$^o$) to $\approx$ 35 \% (for 30$^o$).
The constant term $b$ is  in the range of  $1 \div 3$ \%.
\item
The $e / \pi$-ratios reveal the periodic variation as a function of
$Z$ coordinate with spreads $4 \div 13$~\%.
\item
For some $E,\ \Theta,\ Z$ values the local compensation ($e / h = 1$)
is observed.
\item
Extracted $e / h$ values show the general decreasing tendency
for iron and scintillator thicknesses increasing.
\item
Our results for $e / h$-ratio are compatible with Monte Carlo calculations
of \cite{gabriel} and in disagreement with \cite{wigmans}.
\end{itemize}

\section{Acknowledgements}

This work is the result of the efforts of many people from ATLAS
Collaboration.
The authors are greatly indebted to all Collaboration
for their test beam setup and data taking.

Authors are grateful Peter Jenni and Nikolai Russakovich for their attention
and support of this work.
We are thankful Ana Henriques for her support of this
analysis and constructive advices on the improvement of paper context.
We are indebted to M.~Bosman and M.~Cavalli-Sforza for
the valuable discussions.
%and comments on this note.
We are also thankful to I.~Efthymiopoulos and I.~Chirikov-Zorin for the
%valuable discussions,
careful reading and constructive advices on the improvement
of paper context.

%\newpage

%\newpage

%\listoftables

%\listoffigures

%\newpage

\begin{table}[tbph]
\caption{
        The number of analyzed electron and pion runs as a function
        of incident energy, $\Theta$ and $Z$ coordinate of impact
        on the calorimeter face.
\label{Tb1}}
\begin{center}
\begin{tabular}{|c|c|c|c|c|}
\hline
Energy &${\Theta}^{o}$&Z   &Number   &Number        \\
($GeV$)  &              &(cm)&of $e$ Runs&of $\pi$ Runs \\
\hline
\hline
    & 10 & - 8  $\div$ - 6  & 2 & 2 \\
\cline{2-3}
 20 & 20 & - 20 $\div$ - 18 & 2 & 2 \\
\cline{2-3}
    & 30 & - 38 $\div$ - 36 & 3 & 2 \\
\hline
\hline
    & 10 & - 8  $\div$ - 6   & 2 & 2 \\
\cline{2-3}
 50 & 20 & - 20 $\div$ - 18  & 2 & 2 \\
\cline{2-3}
    & 30 & - 38 $\div$ - 36  & 2 & - \\
\hline
\hline
    & 10 & - 8  $\div$ - 6   & 3 & 2 \\
\cline{2-3}
100 & 20 & - 20 $\div$ - 18  & 2 & 2 \\
\cline{2-3}
    & 30 & - 38 $\div$ - 36  & 3 & - \\
\hline
\hline
    & 10 & - 8  $\div$ - 6   & 2 & 2 \\
\cline{2-3}
150 & 20 & - 20 $\div$ - 18  & 3 & - \\
\cline{2-3}
    & 30 & - 38 $\div$ - 36  & 3 & 2 \\
\hline
\hline
    & 10 & - 8  $\div$ - 6   & 4 & 2 \\
\cline{2-3}
300 & 20 & - 20 $\div$ - 18  & 2 & 2 \\
\cline{2-3}
   & 30 & - 38 $\div$ - 36  & 3 & 1 \\
\hline
\end{tabular}
\end{center}
\end{table}

\begin{table}[tbph]
\caption{
        Cuts for electron events selection.
\label{Tb2}}
\begin{center}
\begin{tabular}{|l|l|l|l|}
\hline
\multicolumn{2}{|l|}{Cut} & min & max \\
\hline
\hline
1 & $Z^{imp}_{beam}$    & $- 14\ mm$ & $8\ mm$ \\
\hline
\hline
2 & $Y^{imp}_{beam}$    & $- 15\ mm$ & $10\ mm$ \\
\hline
\hline
3 & $E_{tot}$    & $E_{tot}^{min}$ & $E_{tot}^{max}$ \\
\hline
\hline
\multicolumn{3}{|l|}{Cut} & min \\
\hline
\hline
4 &\multicolumn{2}{|l|}{$E$ = 20 $GeV$ \& $S_{Cer}$} & 180 \\
\hline
\hline
5 &\multicolumn{2}{|l|}{$E$ $\geq$ 50 $GeV$ \& $\Theta$ = 10$^o$, $C_{i}, i = 3$}
  &$1 - 2 \cdot \sigma$\\
\cline{2-4}
  &\multicolumn{2}{|l|}{$E$ $\geq$ 50 $GeV$ \& $\Theta$ = 20$^o$, $C_{i}, i = 2, 3$}
  &$1 - 2 \cdot \sigma$\\
\cline{2-4}
  &\multicolumn{2}{|l|}{$E$ $\geq$ 50 $GeV$ \& $\Theta$ = 30$^o$, $C_{i}, i = 1, 2$}
  &$1 - 2 \cdot \sigma$\\
\hline
\end{tabular}
\end{center}
\end{table}

%\newpage

\begin{table}[tbph]
\caption{
       The mean relative electron responses $R$ and corresponding
       spreads $\Delta R$ as a function of energy and angle.
\label{Tb3}}
\begin{center}
\begin{tabular}{|c|c|c|c|}
\hline
\multicolumn{4}{|c|}{$R = E / E_{beam}$, ($pC / GeV$)} \\
\hline
\hline
 $E_{beam}$, $GeV$ $\backslash$ ${\Theta}^{o}$ &10$^o$&20$^o$ & 30$^o$ \\
\hline
\hline
20 &   5.64$\pm$0.09 & 5.68$\pm$0.03 & 5.79$\pm$0.03 \\
\hline
50 &   5.42$\pm$0.06 & 5.75$\pm$0.03 & 5.90$\pm$0.03 \\
\hline
100 &  5.52$\pm$0.05 & 5.77$\pm$0.02 & 5.85$\pm$0.02 \\
\hline
150 &  5.40$\pm$0.02 & 5.67$\pm$0.02 & 5.73$\pm$0.02 \\
\hline
300 &  5.78$\pm$0.04 & 5.72$\pm$0.02 & 5.90$\pm$0.02\\
\hline
\hline
$<R_{\Theta}>$ &  5.53$\pm$0.05 & 5.72$\pm$0.06 & 5.83$\pm$0.05\\
\hline
$<\Delta R_{\Theta}>$&$\pm$0.15 (2.8\%)&$\pm$0.16 (2.8\%)&$\pm$0.11 (1.9\%)\\
\hline
$<R>$ & \multicolumn{3}{|c|}{5.69$\pm$0.04$\pm$0.15}\\
\hline
\multicolumn{4}{c}{\mbox{~}}\\[-3mm]
\hline
\multicolumn{4}{|c|}{Spreads $\Delta R$, ($\Delta R / R$)}\\
\hline
\hline
20  & $\pm$0.9 (16.0\%) & $\pm$0.30 (9.4\%) & $\pm$0.25 (4.3\%) \\
\hline
50  & $\pm$0.7 (12.9\%) & $\pm$0.25 (4.4\%) & $\pm$0.20 (3.4\%) \\
\hline
100 & $\pm$0.6 (10.9\%) & $\pm$0.25 (4.4\%) & $\pm$0.20 (3.4\%) \\
\hline
150 & $\pm$0.3 (5.6\%)  & $\pm$0.25 (4.4\%) & $\pm$0.15 (2.6\%) \\
\hline
300 & $\pm$0.35 (6.3\%) & $\pm$0.20 (3.7\%) & $\pm$0.20 (3.5\%) \\
\hline
\end{tabular}
\end{center}
\end{table}

\begin{table}[tbph]
\caption{
        Comparison of measured and calculated electron responses
        for $E = 20\ GeV$ at $10^o$.
\label{Tb05}}
\begin{center}
\begin{tabular}{|c|c|c|c|c|}
\hline
$Z$, mm  & Element & Total $N_e$ & \multicolumn{2}{|c|}{$R,\ (pC / GeV)$} \\
\cline{4-5}
         &         &             & calcul. & exper. \\
\hline
$- 88.5$ & spacer  & 384         & 6.5     & 6.5  \\
\hline
$- 84.0$ & master  & 333         & 5.6     & 5.7  \\
\hline
$- 79.5$ & tile    & 291         & 4.9     & 4.9  \\
\hline
$- 75.0$ & master  & 354         & 5.9     & 5.8  \\
\hline
\end{tabular}
\end{center}
\end{table}

%\newpage

\begin{table}[tbph]
\caption{
        The energy resolution as a function of energy and incidence angles.
\label{Tb5}}
\begin{center}
\begin{tabular}{|c|c|c|c|}
\hline
 $E$, $GeV$ $\backslash$ ${\Theta}^{o}$ &  10$^o$ & 20$^o$ & 30$^o$ \\
\hline
\hline
20 &   14.86$\pm$0.11 & 9.32$\pm$0.15 & 8.89$\pm$0.13 \\
\hline
50 &    9.38$\pm$0.13 & 5.85$\pm$0.10 & 4.43$\pm$0.07 \\
\hline
100 &   6.94$\pm$0.14 & 3.85$\pm$0.05 & 3.36$\pm$0.06 \\
\hline
150 &   5.90$\pm$0.15 & 4.16$\pm$0.15 & 2.21$\pm$0.17 \\
\hline
300 &   4.96$\pm$0.13 & 2.92$\pm$0.10 & 2.44$\pm$0.05 \\
\hline
\end{tabular}
\end{center}
\end{table}

%\newpage

\begin{table}[tbph]
\caption{
        The values of the parameters $a$ and $b$ in the parametrization  of the energy resolution.
\label{Tb6}}
\begin{center}
\begin{tabular}{|c|c|c|c|c|}
\hline
 & \multicolumn{2}{|c|}{\bf $\sigma / E = a / \sqrt{E} + b$}
 & \multicolumn{2}{|c|}{\bf $\sigma / E = a / \sqrt{E} \oplus b$}
\\
\hline
\hline
${\Theta}^{o}$ & $a$ (\%) & $b$ (\%) & $a$ (\%) & $b$ (\%)
\\
\hline
   10 &      61.1$\pm$1.8  & 1.0$\pm$0.3
      &      64.6$\pm$0.7  & 2.9$\pm$0.2
\\
\hline
   20 &      38.6$\pm$3.0  & 0.3$\pm$0.3
      &      39.1$\pm$2.6  & 1.3$\pm$0.9
\\
\hline
   30 &      33.4$\pm$4.9  & 0.2$\pm$0.5
      &      33.7$\pm$1.1  & 1.1$\pm$1.1
\\
\hline
\end{tabular}
\end{center}
\end{table}

%\newpage

\begin{table}[tbph]
\caption{
        The values of parameter $a$ for existing experimental data for
        iron-scintillator calorimeters.
\label{Tb11}}
\begin{center}
\begin{tabular}{|l|l|c|c|c|c|c|}
\hline
Author     & Ref.             &$t$&$s$& $a$          & ${a}_{calc}$\\
\hline
\hline
Stone      & \cite{stone}     &4.8&6.3& 10.          & 7.0            \\
\hline
Antipov    & \cite{antipov}   &20.&5.0& 27.          & 17.               \\
\hline
Abramovicz & \cite{abram}     &25.&5.0& 23.          & 20.             \\
\hline
our data 30$^o$     &         &28.&6.0& $34. \pm 1.$ & 20.             \\
\hline
our data 20$^o$     &         &41.&9.0& $39. \pm 3.$ & 24.             \\
\hline
our data 10$^o$     &         &81.&17.& $65. \pm 1.$ & 32.             \\
\hline
\end{tabular}
\end{center}
\end{table}

%\newpage

\begin{table}[tbph]
\caption{
       The mean $e/ \pi$ ratios and the corresponding spreads as
        a function of the incidence energy and angle.
\label{Tb7}}
\begin{center}
\begin{tabular}{|c|c|c|c|}
\hline
\multicolumn{4}{|c|}{$e/ \pi$ ratio}\\
\hline
\hline
$E$, $GeV$ $\backslash$ ${\Theta}^{o}$ &  10$^o$ & 20$^o$ & 30$^o$ \\
\hline
\hline
20 &    1.148$\pm$0.017 & 1.178$\pm$0.006 & 1.197$\pm$0.004 \\
\hline
50 &    1.096$\pm$0.012 & 1.172$\pm$0.005 & 1.198$\pm$0.006 \\
\hline
100 &   1.102$\pm$0.009 & 1.155$\pm$0.004 & 1.182$\pm$0.005 \\
\hline
150 &   1.083$\pm$0.008 & 1.141$\pm$0.005 & 1.165$\pm$0.004 \\
\hline
300 &   1.088$\pm$0.006 & 1.106$\pm$0.004 & 1.149$\pm$0.005 \\
\hline
\hline
\multicolumn{4}{|c|}{Spreads for $e/ \pi$ ratio}\\
\hline
\hline
20   & $\pm$0.15 (13.1\%) & $\pm$0.063 (5.4\%) & $\pm$0.055 (4.6\%) \\
\hline
50   & $\pm$0.14 (12.8\%) & $\pm$0.050 (4.3\%) & $\pm$0.065 (5.4\%) \\
\hline
100  & $\pm$0.11 (10.0\%) & $\pm$0.050 (4.3\%) & $\pm$0.055 (4.7\%) \\
\hline
150  & $\pm$0.063 (5.8\%) & $\pm$0.080 (7.1\%) & $\pm$0.060 (5.2\%) \\
\hline
300  & $\pm$0.075 (7.2\%) & $\pm$0.038 (3.5\%) & $\pm$0.050 (4.5\%) \\
\hline
\end{tabular}
\end{center}
\end{table}

%\newpage

\begin{table}[tbph]
\caption{
        The $e / h$ ratios.
\label{Tb8}}
\begin{center}
\begin{tabular}{|l|l|c|c|c|c|c|}
\hline
Author     & Ref.                &$R_d$&$t$, mm&$s$, mm&$e / h$ & Symbols \\
\hline
\hline
Bohmer  & \cite{bohmer}           &2.8  &20.    &7.0    &1.44   &
{\Large $\bullet$} \\
\hline
Wigmans     & \cite{wigmans}~$^*$ &3.0  &15.    &5.0    &1.25   &
$\blacktriangle$ \\
\hline
Antipov     & \cite{antipov}      &4.0  &20.    &5.0    &1.15   &
{\Large $\circ$} \\
\hline
Wigmans     & \cite{wigmans}~$^*$ &4.0  &20.    &5.0    &1.23   &
$\blacktriangle$ \\
\hline
TileCal, 30$^o$   &               &4.7  &28.    &6.0    &$1.39 \pm 0.03$ &
$\blacksquare$ \\
\hline
TileCal,  20$^o$& \cite{ariz-94}&4.7  &41.    &9.0    &$1.37 \pm 0.02$ &
$\square$ \\
\hline
TileCal, 20$^o$   &               &4.7  &41.    &9.0    &$1.34 \pm 0.03$ &
$\blacksquare$ \\
\hline
TileCal, 10$^o$   &               &4.7  &81.    &17.    &$1.23 \pm 0.02$ &
$\blacksquare$ \\
\hline
Wigmans     & \cite{wigmans}~$^*$ &5.0  &25.    &5.0    &1.21   &
$\blacktriangle$      \\
\hline
Abramovicz  & \cite{abram}        &5.0  &25.    &5.0    &1.32   &
$\lozenge$  \\
\hline
Vincenzi    & \cite{vince}        &5.0  &25.    &5.0    &1.32   &
$\bigstar$ \\
\hline
Wigmans     & \cite{wigmans}~$^*$ &6.0  &30.    &5.0    &1.20   &
$\blacktriangle$      \\
\hline
Gabriel     & \cite{gabriel}~$^*$ &6.3  &19.    &3.0    &1.55   &
$\blacktriangledown$       \\
\hline
Wigmans     & \cite{wigmans}~$^*$ &8.0  &40.    &5.0    &1.18   &
$\blacktriangle$      \\
\hline
Holder      & \cite{holder}       &8.3  &50.    &6.0    &1.18   &
{\Large $\ast$}  \\
\hline
Gabriel     & \cite{gabriel}~$^*$ &8.5  &25.4   &3.0    &1.50   &
$\blacktriangledown$      \\
\hline
Wigmans     & \cite{wigmans}~$^*$ &10.  &50.    &5.0    &1.16   &
$\blacktriangle$      \\
\hline
\multicolumn{7}{l}{}\\[-2mm]
\cline{1-3}
\multicolumn{7}{l}{}\\[-4mm]
\multicolumn{7}{l}{$^*$~Monte Carlo calculations}\\
\end{tabular}
\end{center}
\end{table}

%\end{document}

\newpage

\begin{figure*}[tbph]
     \begin{center}
       %\vspace*{1in}
      \mbox{\epsfig{figure=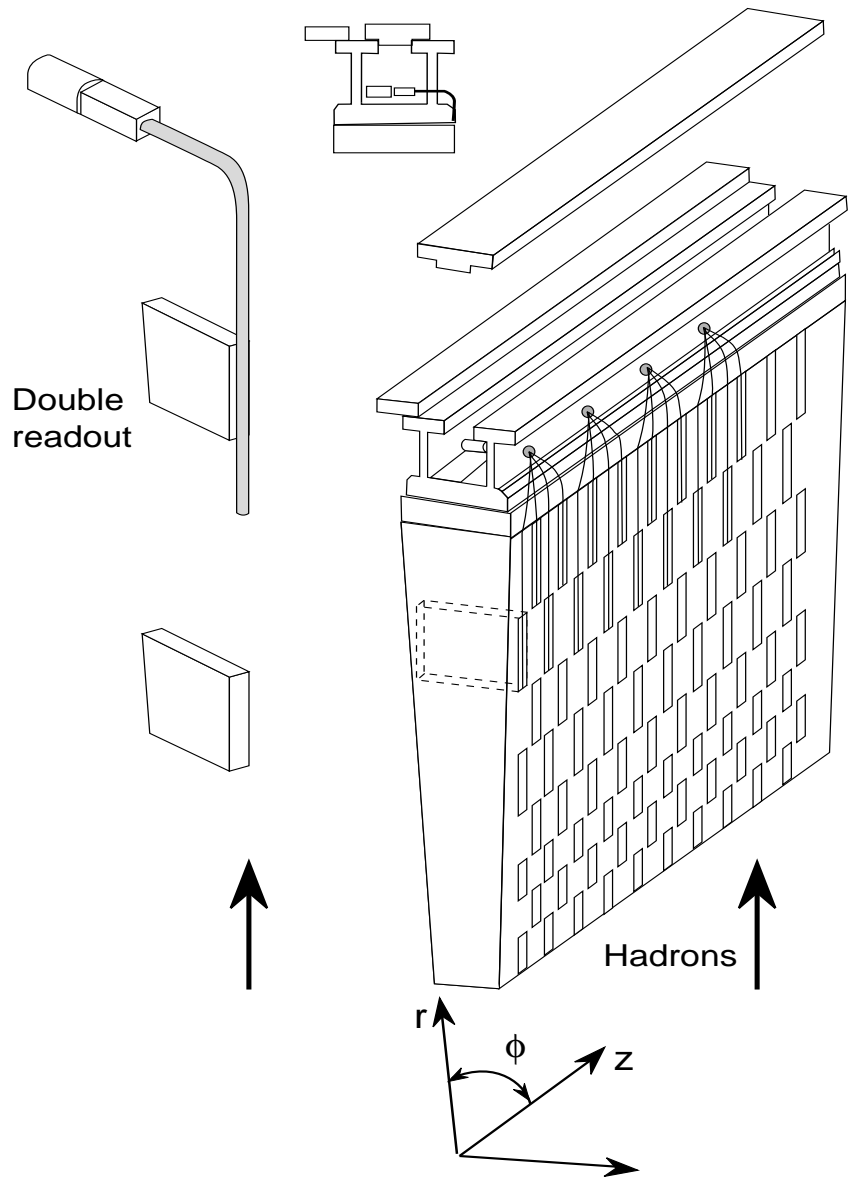,width=0.9\textwidth,height=0.8\textheight}}
     \end{center}
       \caption{
       Principal of the tile hadronic calorimeter.
       \label{f2-28tp}}
\end{figure*}

%\newpage

\begin{figure*}[tbph]
     \begin{center}
       \mbox{\epsfig{figure=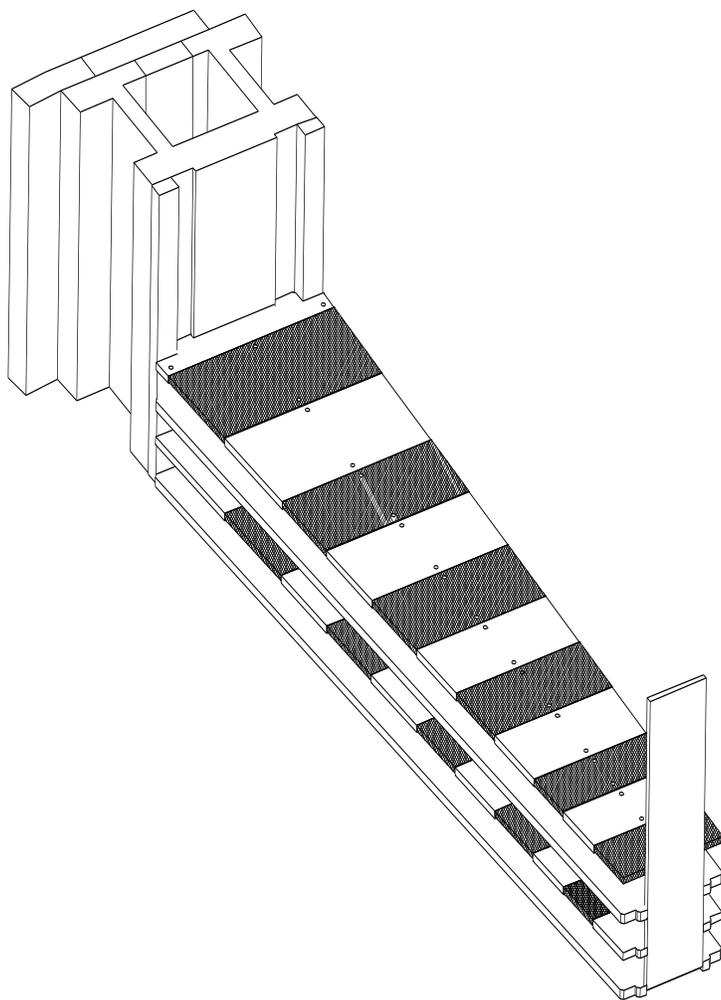,width=0.95\textwidth,height=0.6\textheight}}
       \vspace{10mm}
    \end{center}
       \caption{
        Exploded view of an assembled tile calorimeter period.
       \label{f2-29tp}}
\end{figure*}

%\end{document}

%\newpage

\begin{figure*}[tbph]
     \begin{center}
       \mbox{\epsfig{figure=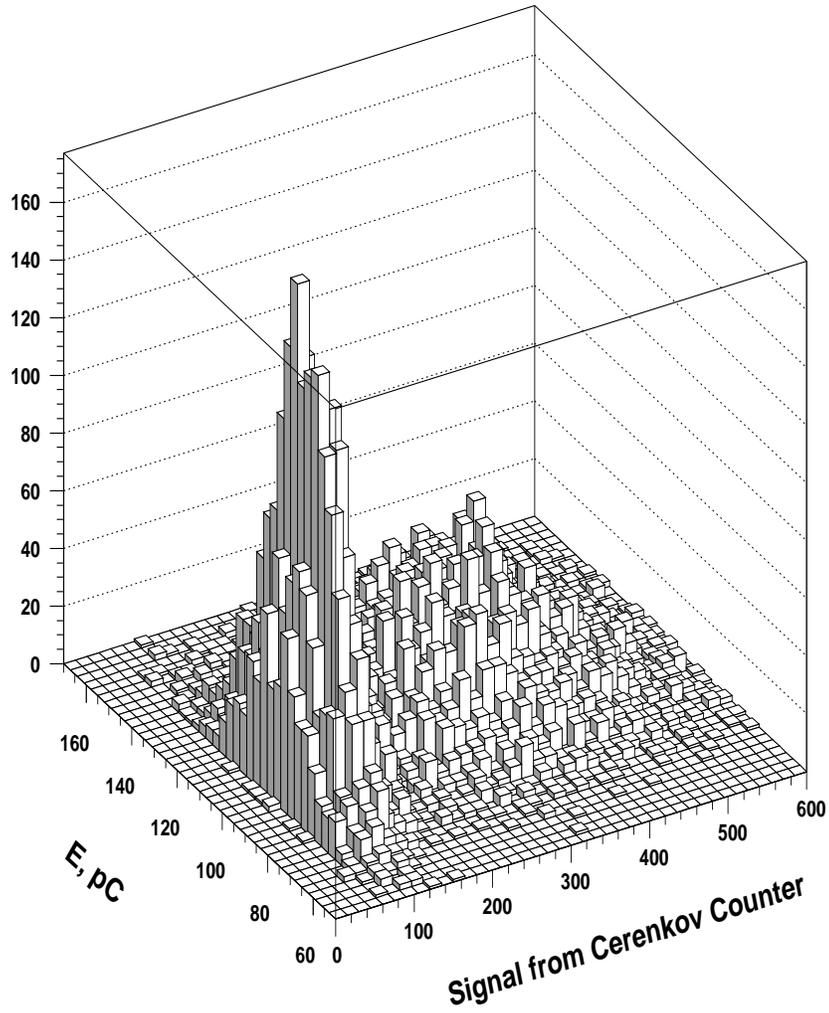,width=0.95\textwidth,height=0.8\textheight}}
     \end{center}
       \caption{
        Distribution of the events for $E = 20\ GeV$, $\Theta = 10^{o}$ as a
       function of values $E_{tot}$ and \v{C}erenkov counter signal.
       \label{fig:f4}}
\end{figure*}

%\newpage

\begin{figure*}[t]
     \begin{center}
      \mbox{\epsfig{figure=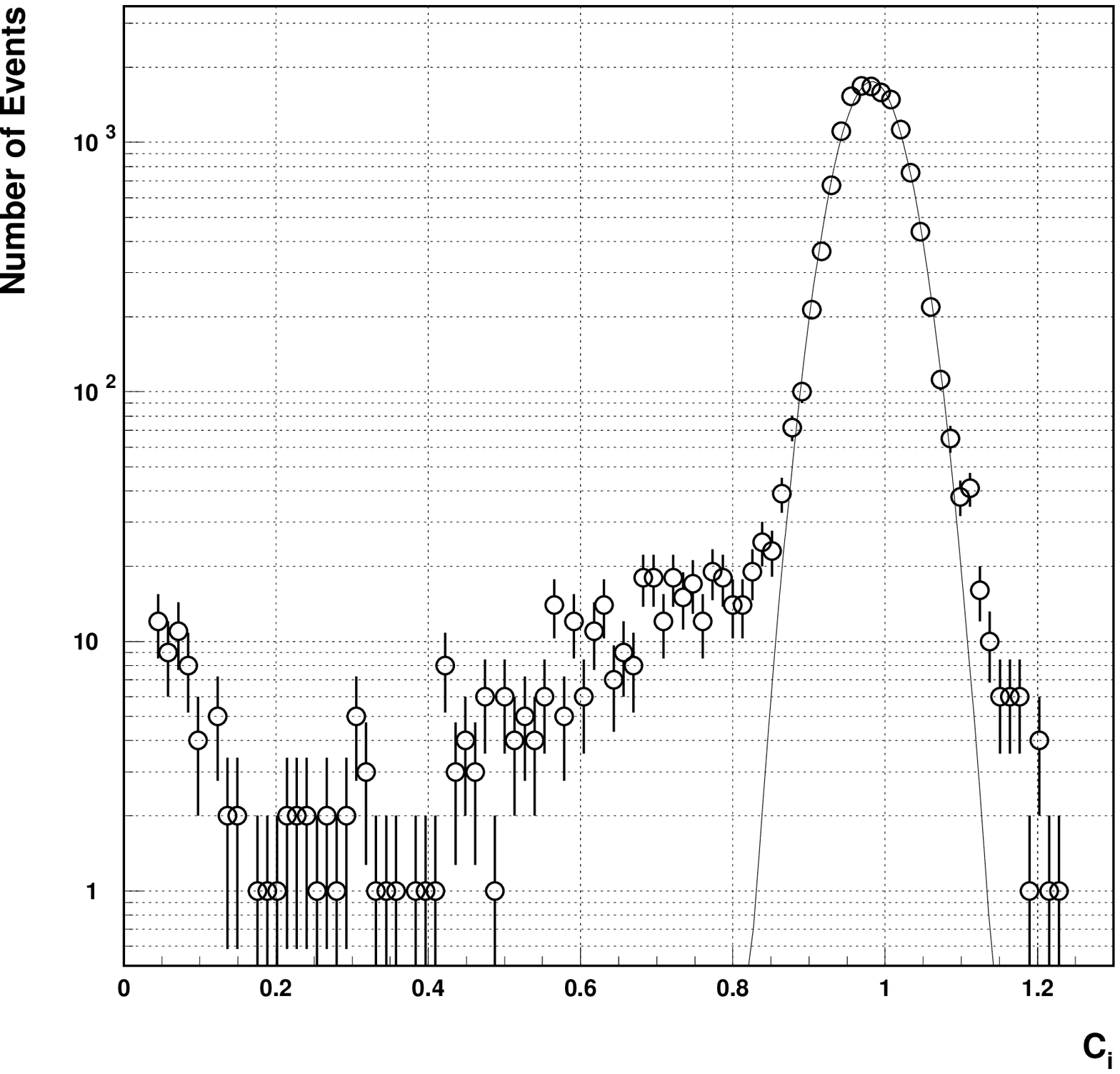,width=0.95\textwidth,height=0.8\textheight}}
     \end{center}
      \caption{
        Distribution of the events for $E = 20\ GeV$, $\Theta = 10^{o}$ as a
        function of $C_i$ for electrons tagged by \v{C}erenkov counter.
       \label{fig:f5a}}
\end{figure*}

\begin{figure*}[t]
     \begin{center}
      \mbox{\epsfig{figure=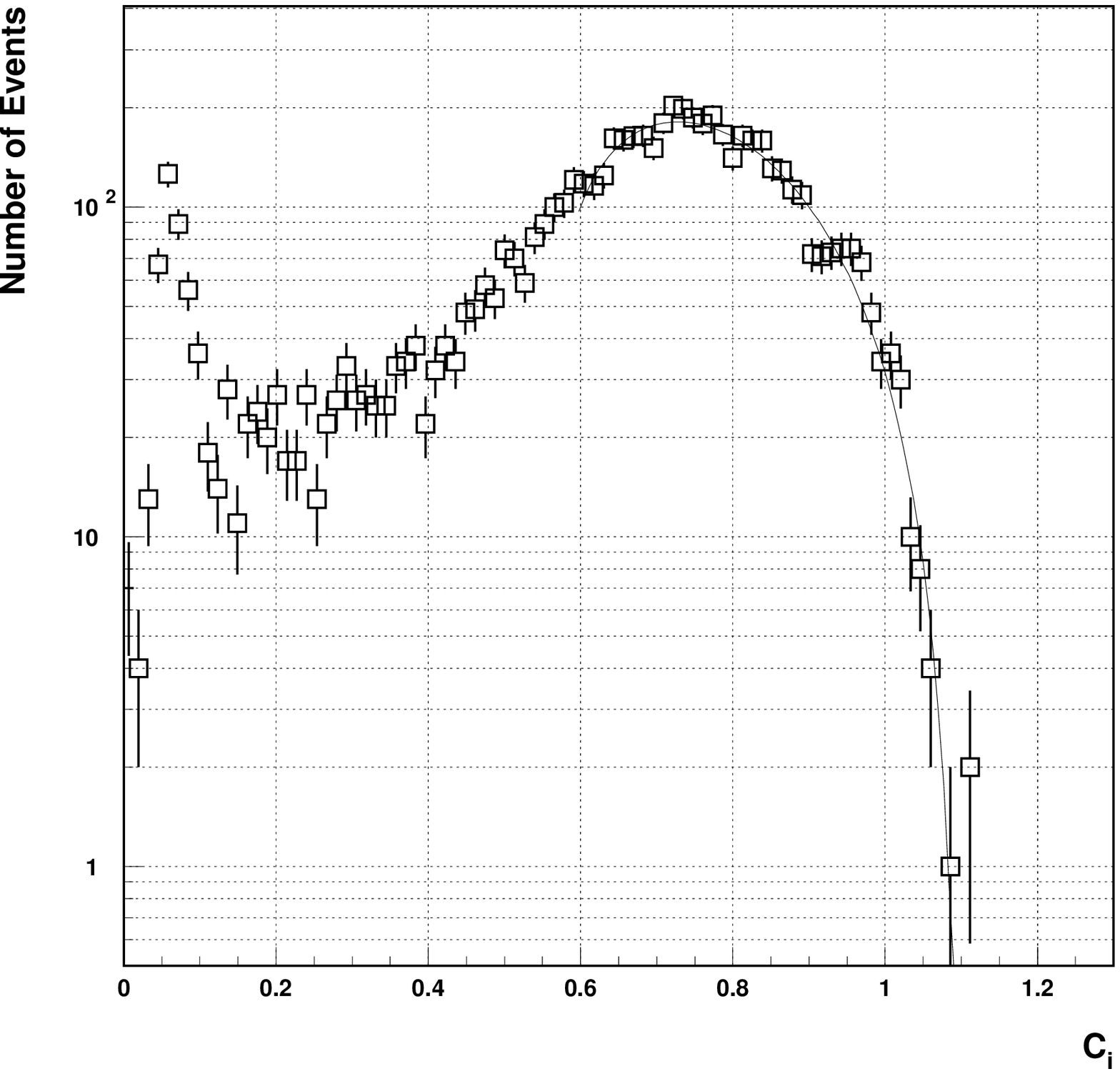,width=0.95\textwidth,height=0.8\textheight}}
     \end{center}
      \caption{
        Distribution of the events for $E = 20\ GeV$, $\Theta = 10^{o}$ as a
        function of $C_i$ for $\pi$- mesons tagged by \v{C}erenkov counter.
       \label{fig:f5b}}
\end{figure*}

\begin{figure*}[b]
     \begin{center}
      \mbox{\epsfig{figure=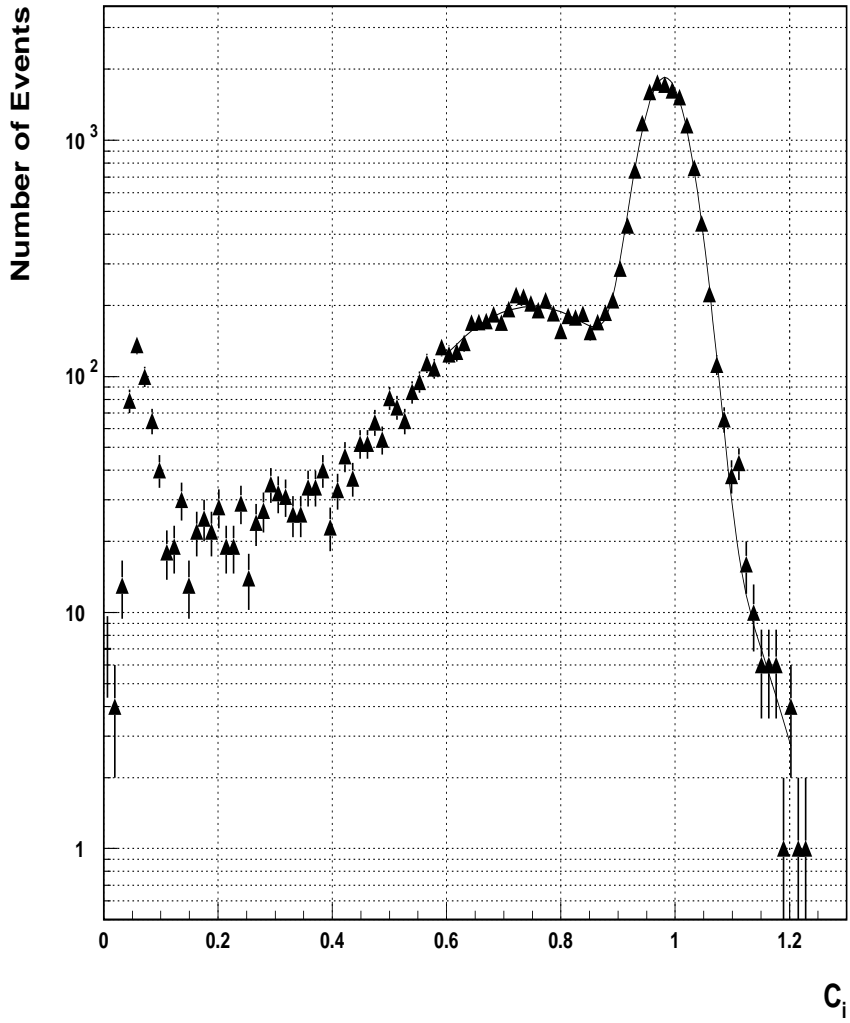,width=0.95\textwidth,height=0.8\textheight}}
     \end{center}
      \caption{
        Distribution of the events for $E = 20\ GeV$, $\Theta = 10^{o}$ as a
        function of $C_i$ for all events.
       \label{fig:f5c}}
\end{figure*}

%\newpage

\begin{figure*}[tbph]
     \begin{center}
       \mbox{\epsfig{figure=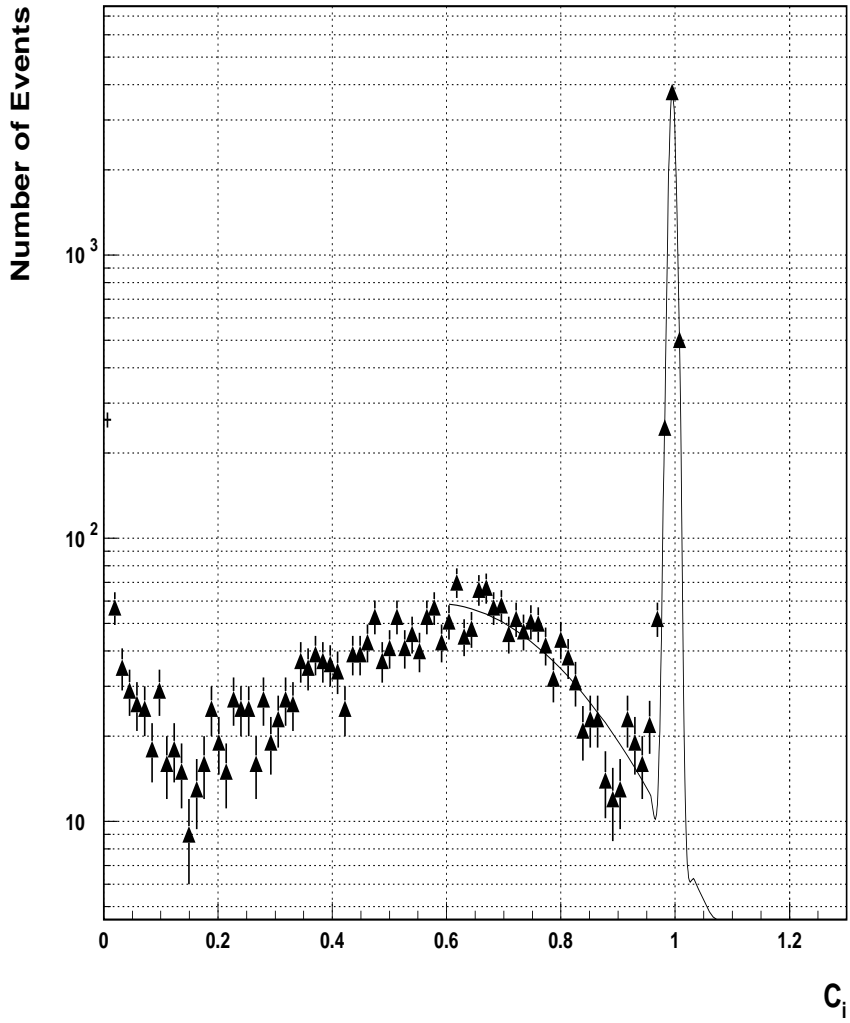,width=0.95\textwidth,height=0.8\textheight}}
     \end{center}
       \caption{
        Distribution of the events for $E = 300\ GeV$ electron at $\Theta = 30^{o}$ as a
        function of $C_i$ for all events.
       \label{fig:f6}}
\end{figure*}

%\newpage

\begin{figure*}[tbph]
     \begin{center}
      \mbox{\epsfig{figure=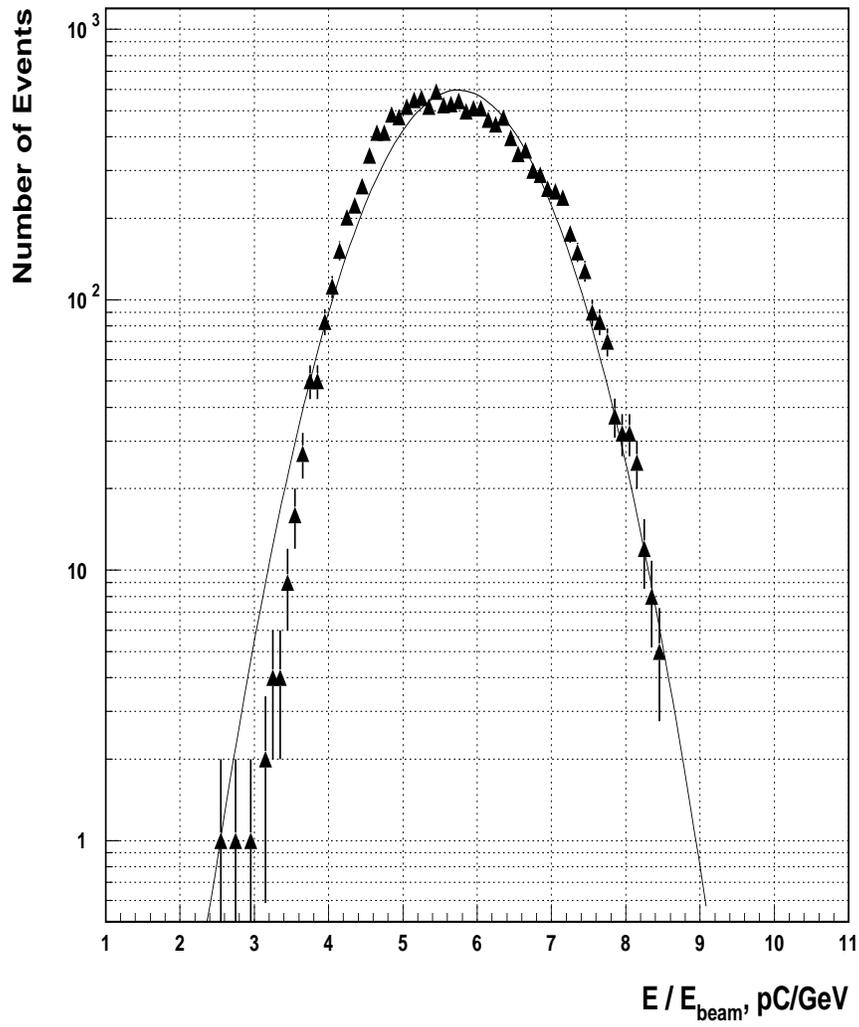,width=0.95\textwidth,height=0.8\textheight}}
     \end{center}
       \caption{
        Energy distribution for 20 $GeV$ electrons at $10^{o}$
        as a function of $Z$ coordinate.
       \label{fig:f07}}
\end{figure*}

%\newpage

\begin{figure*}[tbph]
     \begin{center}
      \mbox{\epsfig{figure=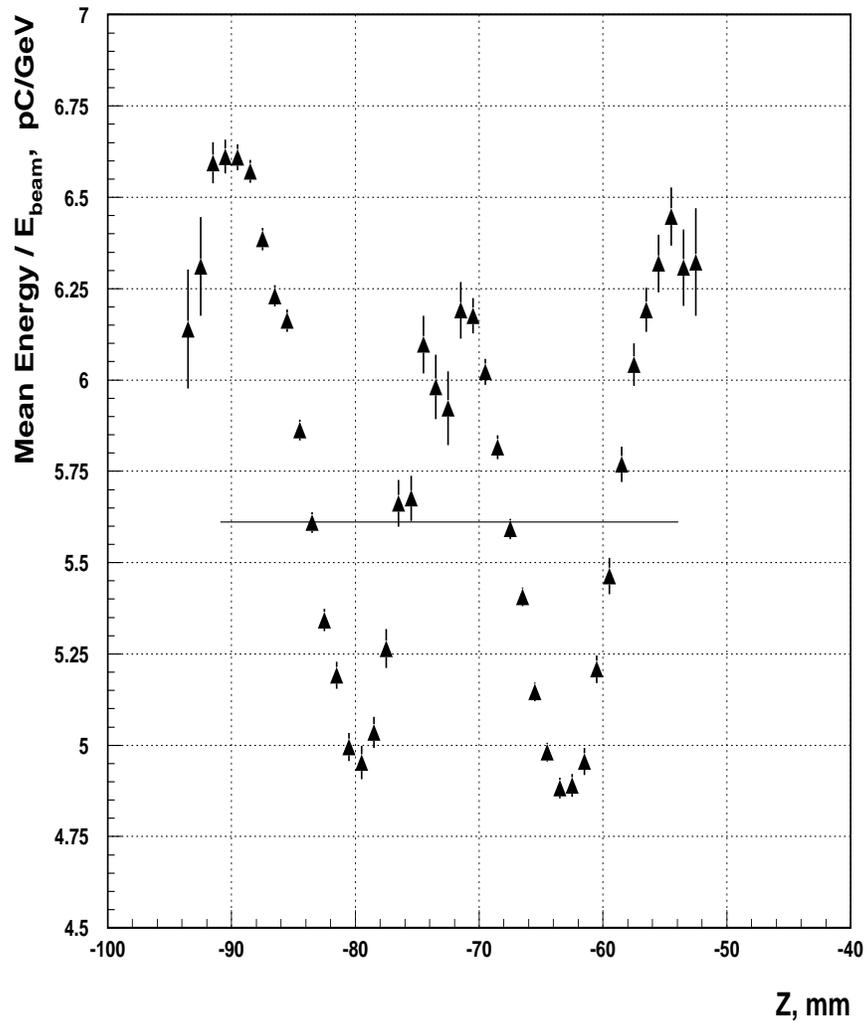,width=0.95\textwidth,height=0.8\textheight}}
     \end{center}
       \caption{
        Response to 20 $GeV$ electrons at $10^{o}$
        as a function of $Z$ coordinate.
       \label{fig:f7}}
\end{figure*}

%\newpage

\begin{figure*}[tbph]
     \begin{center}
       \mbox{\epsfig{figure=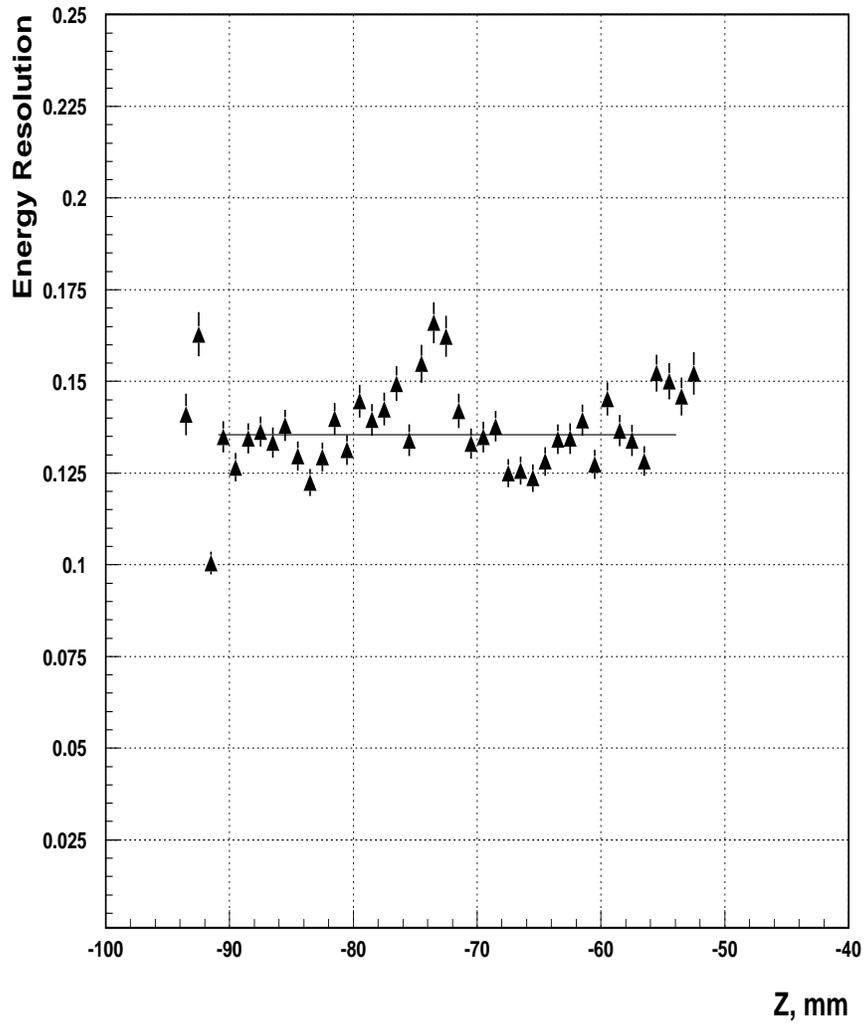,width=0.95\textwidth,height=0.8\textheight}}
     \end{center}
       \caption{
        Energy resolution for 20 $GeV$ electrons at
        $10^{o}$ as a function of $Z$ coordinate.
       \label{fig:f11}}
\end{figure*}

%\newpage

\begin{figure*}[tbph]
     \begin{center}
       \mbox{\epsfig{figure=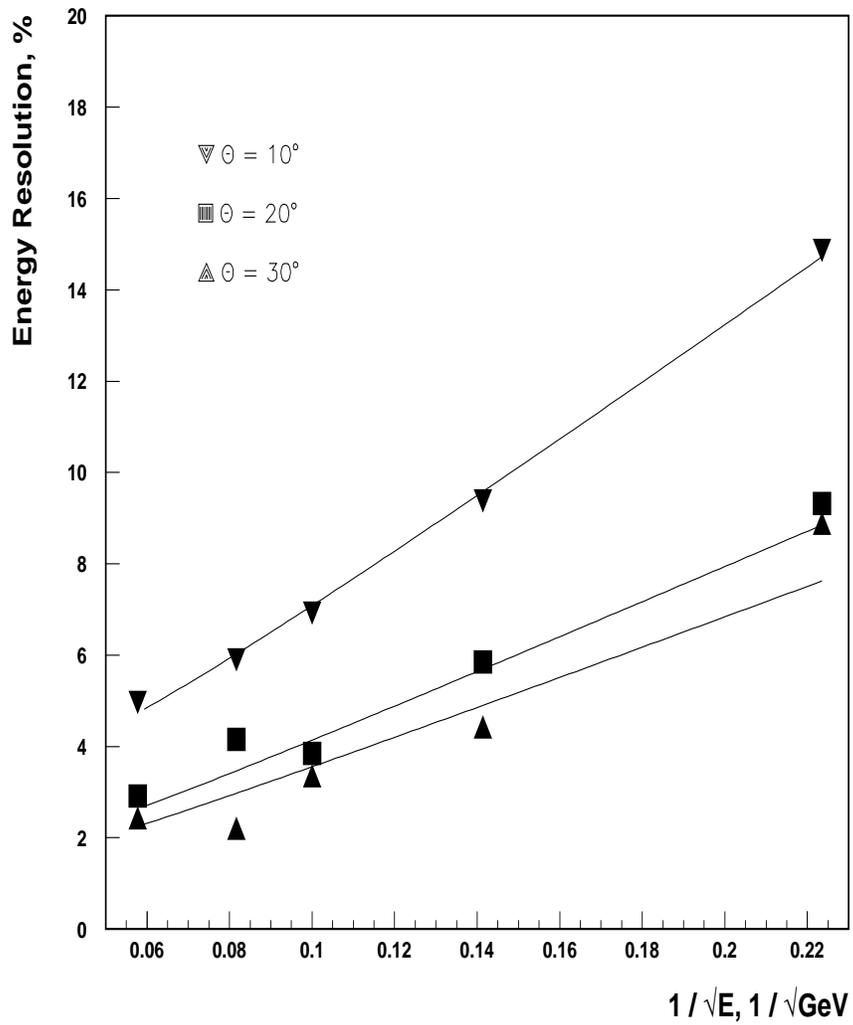,width=0.95\textwidth,height=0.8\textheight}}
    \end{center}
       \caption{
       Energy resolution for electrons at $10^{o}$, $20^{o}$
        and $30^{o}$.
       \label{fig:f12}}
\end{figure*}

%\newpage

\begin{figure*}[tbph]
    \begin{center}
       \mbox{\epsfig{figure=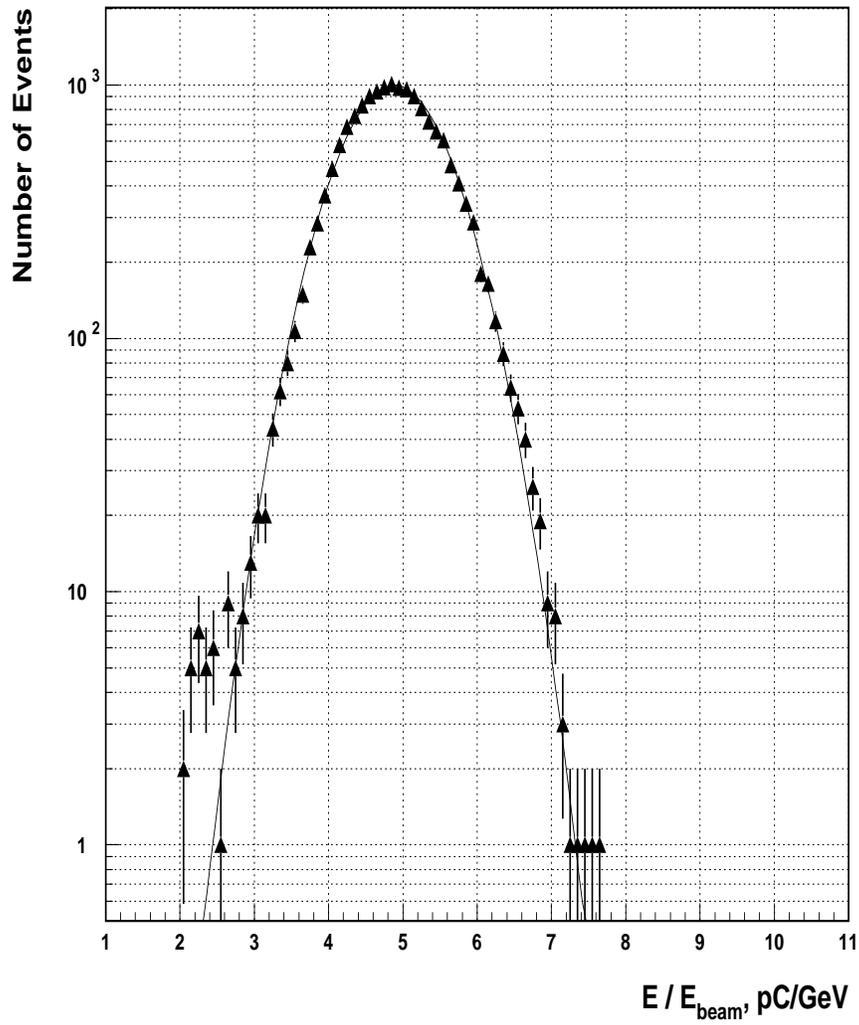,width=0.95\textwidth,height=0.8\textheight}}
     \end{center}
       \caption{
        Energy response for  $20\ GeV$ pions at $20^{o}$.
       \label{fig:f22}}
\end{figure*}

\begin{figure*}[tbph]
    \begin{center}
       \mbox{\epsfig{figure=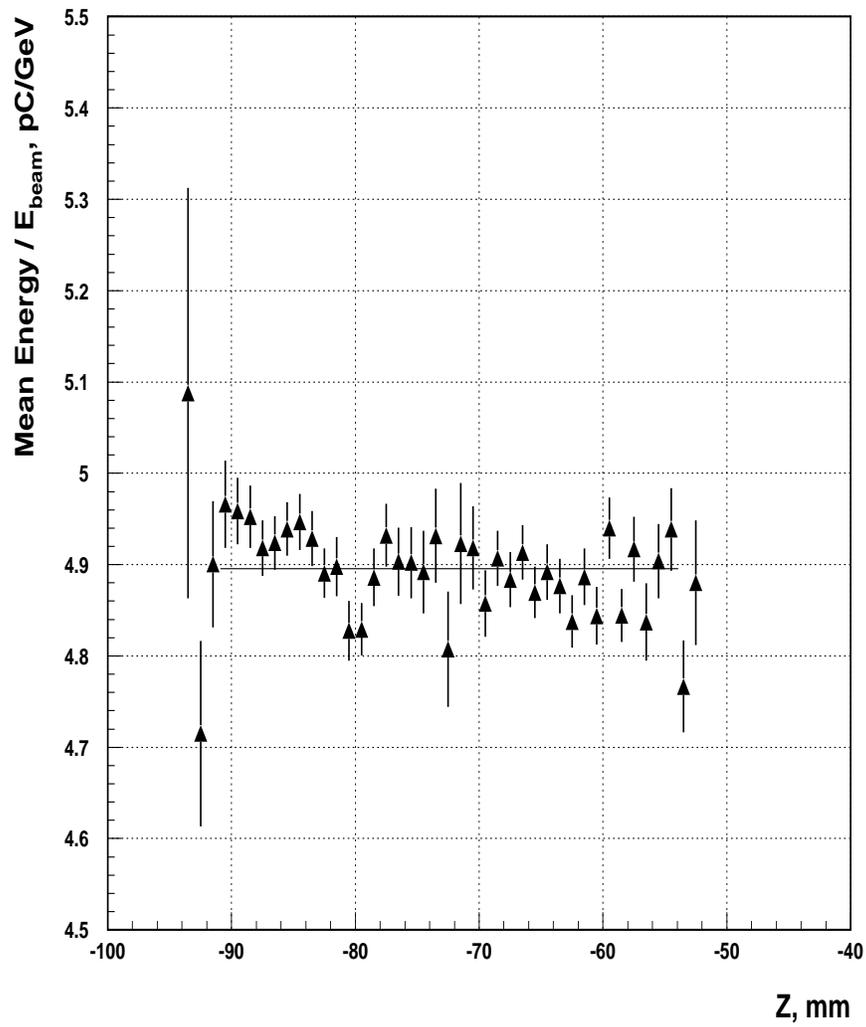,width=0.95\textwidth,height=0.8\textheight}}
     \end{center}
       \caption{
        Pion response for  $20\ GeV$ and $10^{o}$
        as a function of $Z$ coordinate.
       \label{fig:f21}}
\end{figure*}

%\newpage

\begin{figure*}[tbph]
     \begin{center}
       \mbox{\epsfig{figure=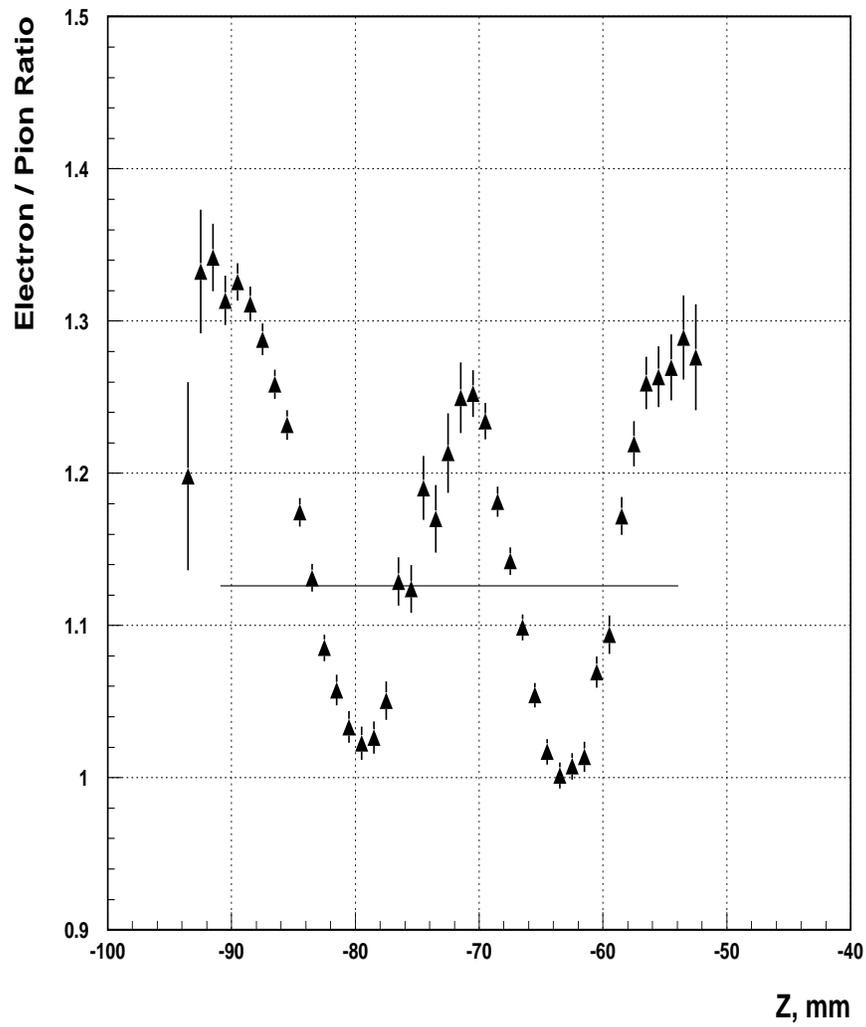,width=0.95\textwidth,height=0.8\textheight}}
     \end{center}
       \caption{
       The $e / \pi$ ratio for 20 $GeV$ at $10^{o}$
        as a function of $Z$ coordinate.
       \label{fig:f13}}
\end{figure*}

%\newpage

\begin{figure*}[tbph]
    \begin{center}
       \mbox{\epsfig{figure=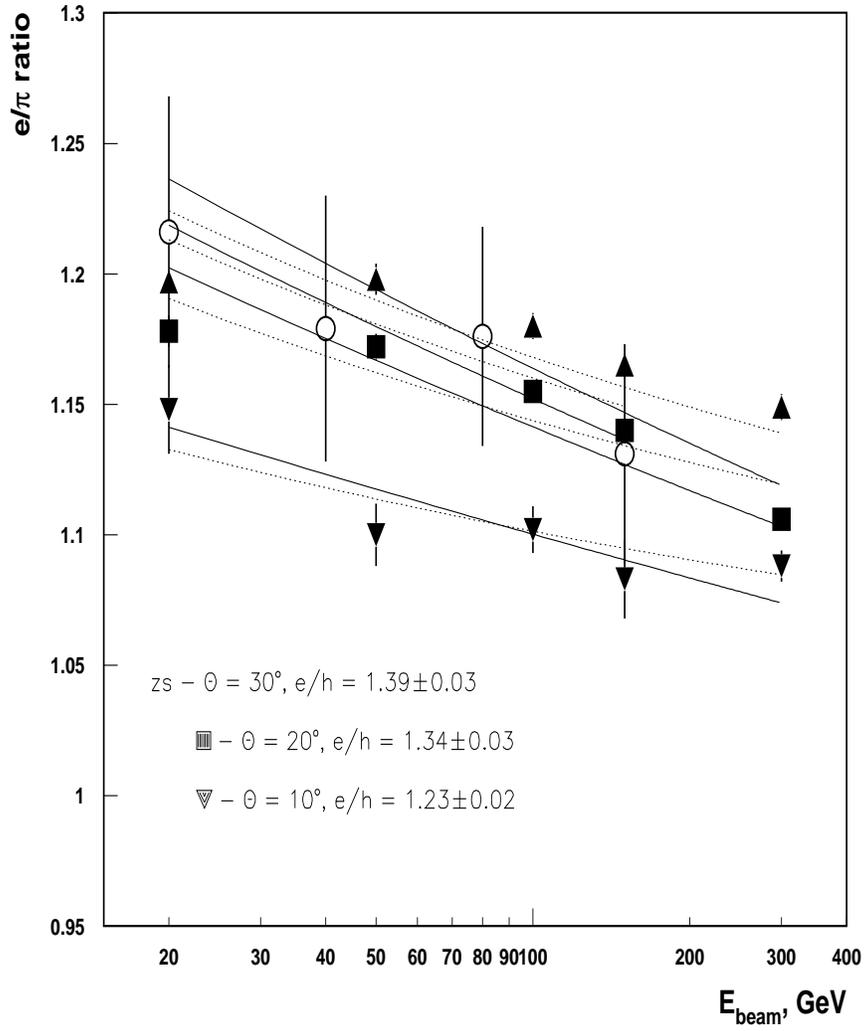,width=0.95\textwidth,height=0.8\textheight}}
     \end{center}
       \caption{
        The $e/ \pi$ ratio as a function of energy at $10^{o}$, $20^{o}$ and
        $30^{o}$.
       \label{fig:f14}}
\end{figure*}

%\newpage

\begin{figure*}[tbph]
     \begin{center}
     \mbox{\epsfig{figure=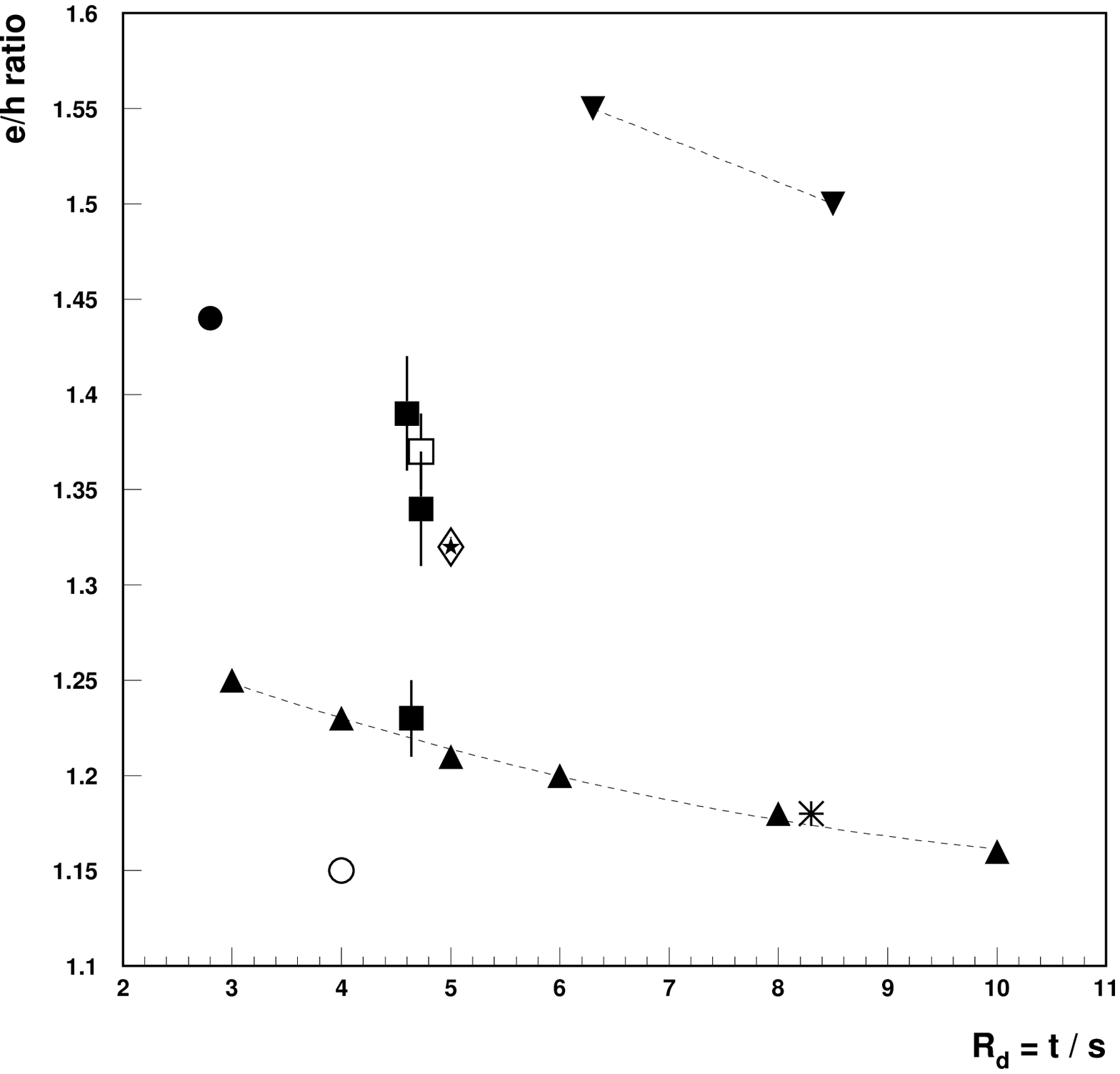,width=0.95\textwidth,height=0.8\textheight}}
     \end{center}
     \caption{
       The $e/ h$-ratios as a function of iron and scintillator thicknesses ratio.
        The lines are the results of fits to some selected data.
        The meaning of symbols see in Table~9.
       \label{fig:f17}}
\end{figure*}

%\newpage

\begin{figure*}[tbph]
     \begin{center}
     \mbox{\epsfig{figure=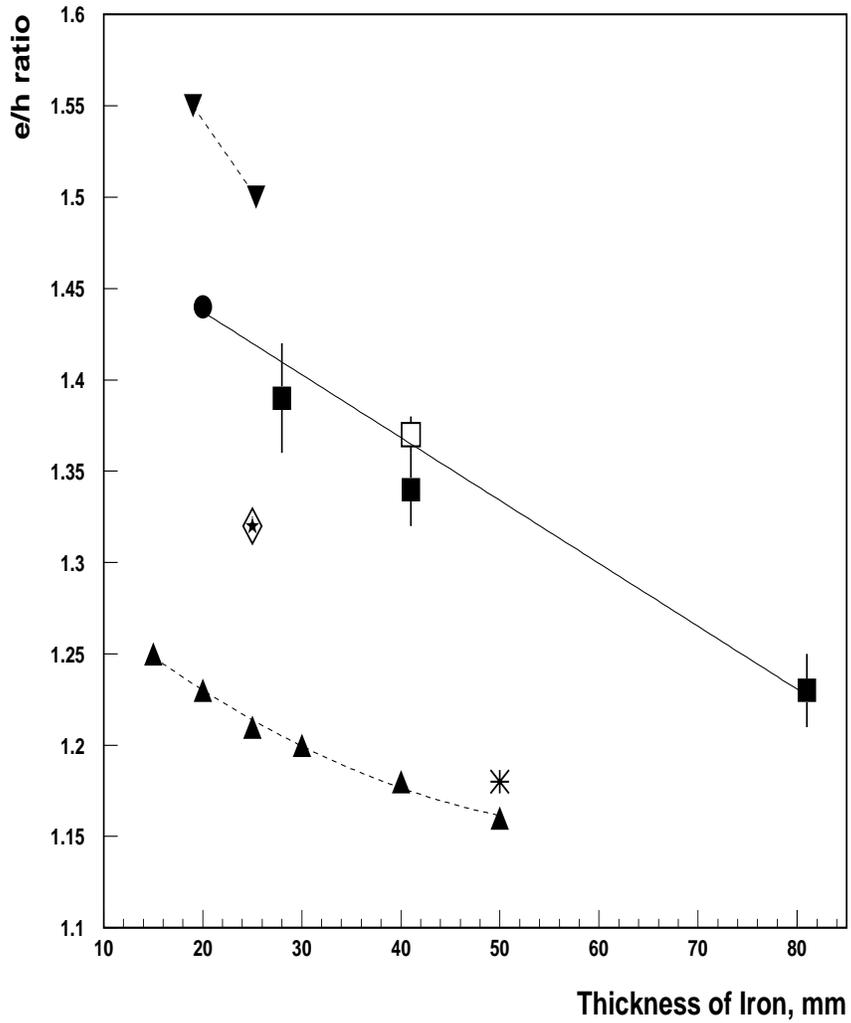,width=0.95\textwidth,height=0.8\textheight}}
     \end{center}
     \caption{
        The $e/ h$-ratios as a function of iron thickness.
        The lines are the results of fits to some selected data.
        The meaning of symbols see in Table~9.
       \label{fig:f15}}
\end{figure*}

%\newpage

\begin{figure*}[tbph]
     \begin{center}
     \mbox{\epsfig{figure=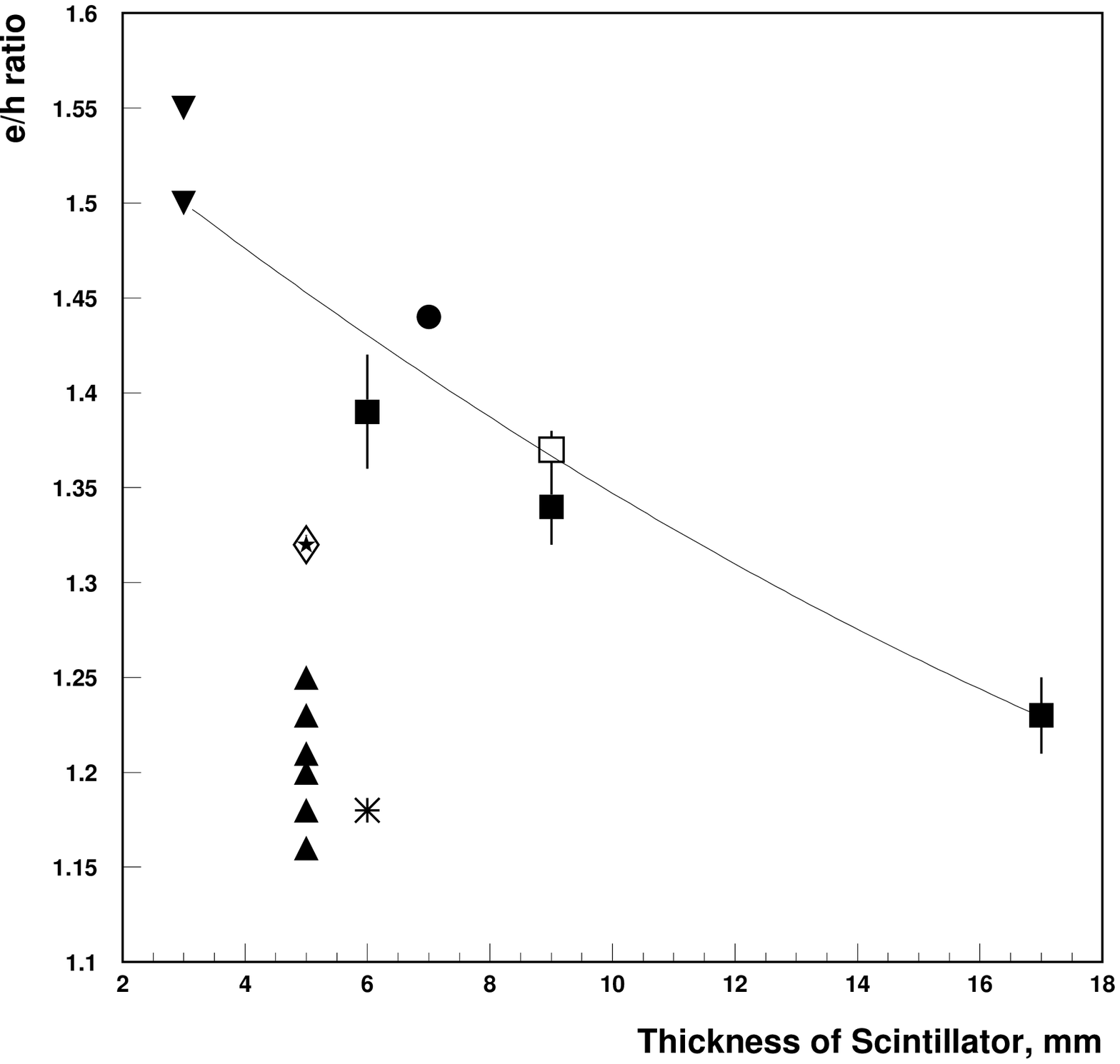,width=0.95\textwidth,height=0.8\textheight}}
     \end{center}
     \caption{
        The $e/ h$-ratios as a function of scintillator thickness.
        The lines is the results of fits to some selected data.
        The meaning of symbols see in Table~9.
       \label{fig:f16}}
\end{figure*}

}

\begin{thebibliography}{00}
%1
\bibitem{atcol}
ATLAS Collaboration, CERN/LHCC/94-93, ATLAS Technical Proposal for a
General-Purpose pp experiment at the Large Hadron Collider CERN,
CERN, Geneva, Switzerland.
%2
\bibitem{lhcnews}
LHC News, $N^{\b{o}}$ 7 September 1995, CERN, Geneva, Switzerland.
%3
\bibitem{gild-91}
O.~Gildemeister, F.~Nessi-Tedaldi and M.~Nessi,
Proc.~2nd Int. Conf. on Cal.~in HEP, Capri, 1991.
%4
\bibitem{ariz-94}
F.~Ariztizabal et al., NIM A349 (1994) 384.
%5
\bibitem{berger}
E.~Berger et al. CERN/LHCC 95-44,
Construction and Performance of an Iron-Scintillator
Hadron Calorimeter with Longitudinal Tile Configuration,
CERN, Geneva, Switzerland.
%6
\bibitem{david}
A.~Amorim, M.~David, ATLAS TILECAL-TR-24,
November 1994, CERN, Geneva, Switzerland.
%6-1
\bibitem{efthym}
M.~Cavalli-Sforza, M.~Pilar~Casado, ATLAS TILECAL-TR-43,
September 1995, CERN, Geneva, Switzerland.
%7
\bibitem{atmon}
I.~Efthymiopoulos, A.~Solodkov,
The TILECAL Program for Test Beam Data Analysis, User Manual, 1995,
CERN, Geneva, Switzerland.
%8
\bibitem{amaldi}
U.~Amaldi, Phys.~Scripta 23 (1981) 409.
%8-1
\bibitem{henriq}
A.~Henriques, G.~Karapetian, A.~Solodkov, ATLAS Internal Note,
TILECAL-No-68, November 1995, CERN, Geneva, Switzerland.
%9
\bibitem{abshire}
G.~Abshire et al., NIM 164 (1979) 67.
%10
\bibitem{prabha}
A.~Prabhaharan and L.~Bugge, NIM A314 (1992) 21.
%11
\bibitem{delpe}
J.~Del~Peso, E.~Ros, NIM A276 (1989) 456.
%12
\bibitem{bosman}
M.~Bosman, ATLAS TILECAL-TR-43, September 1995,
CERN, Geneva, Switzerland.
%13
\bibitem{rwig}
R.~Wigmans, NIM A265 (1988) 273.
%14
\bibitem{wigmans}
R.~Wigmans, NIM A259 (1987) 389.
%15
\bibitem{stone}
S.~L.~Stone et al., NIM 151 (1978) 387.
%16
\bibitem{antipov}
Y.~A.~Antipov et al., NIM 180 (1990) 81.
%17
\bibitem{abram}
H.~Abramowicz et al., NIM 180 (1981) 429.
%18
\bibitem{bohmer}
V.~Bohmer et al., NIM 122 (1974) 313.
%19
\bibitem{holder}
M.~Holder et al., NIM 151 (1978) 69.
%20
\bibitem{vince}
M.~De~Vincenze et al., NIM A243 (1986) 348.
%21
\bibitem{gabriel}
T.~A.~Gabriel et al., NIM A295 (1994) 336.
\end{thebibliography}
\end{document}